# Large magnetic circular dichroism in resonant inelastic x-ray scattering at the Mn *L*-edge of Mn-Zn ferrite

M. Magnuson[1], L.-C. Duda[1], S. M. Butorin[1], P. Kuiper[2] and J. Nordgren[1]

[1]*Department of Physics, Uppsala University, Ångström Laboratory, Box 530, S-75121 Uppsala, Sweden*

[2]*Department of Physics, Växjö University, Vejdes plats 6, S-351 95 Växjö, Sweden*

**Abstract**

We report resonant inelastic x-ray scattering (RIXS) excited by circularly polarized x-rays on Mn-Zn ferrite at the Mn $L_{2,3}$-resonances. We demonstrate that crystal field excitations, as expected for localized systems, dominate the RIXS spectra and thus their dichroic asymmetry cannot be interpreted in terms of spin-resolved partial density of states, which has been the standard approach for RIXS dichroism. We observe large dichroic RIXS at the $L_2$-resonance which we attribute to the absence of metallic core hole screening in the insulating Mn-ferrite. On the other hand, reduced $L_3$-RIXS dichroism is interpreted as an effect of longer scattering time that enables spin-lattice core hole relaxation via magnons and phonons occurring on a femtosecond time scale.

The prediction of magnetic circular dichroism (MCD) in x-ray emission at the *L*-edge of ferromagnetic 3*d*-metals such as iron[1] has triggered much experimental effort to study magnetic effects in x-ray fluorescence spectroscopy[2, 3, 4, 5, 6]. Initial interpretations[1] of MCD in x-ray fluorescence centered around the notion that the MCD spectra of *itinerant* ferromagnetic metals reflect the occupied partial (e.g. Fe 3*d*) density of states (pDOS) and thus treated as a bulk-sensitive complement to spin-resolved photoelectron spectroscopy. On the other hand, one quickly realized that the observed dichroic asymmetries i.e., the relative difference in magnetization specific spectra, are an order of magnitude smaller than theoretically expected. This is puzzling since the initial spin-polarization of the core hole [7, 8, 9] produced by the interaction of the circularly polarized x-rays removing an electron from a 2*p* core level should match the spin polarization of the outgoing photoelectron. Moreover, other issues, such as dichroic saturation or dichroic self-absorption that arise at the magnetic transition metal *L*-edges have hampered development. Instead, much work has been devoted to studying competing atomic like $L_l$(2*p*−3*s*)-decay or using other experimental geometries [10, 11].

Only a few attempts have been made to explain the intriguing blatant discrepancy [12, 13] in the magnitude of theoretically expected and experimentally observed dichroic asymmetries in x-ray emission. It has been shown that it is important to take into account the spin-orbit interaction and the fact that spin is no longer a good quantum number for the core-excited 2*p* level. However, this effect is too small to account for the entire reduction of the dichroic asymmetry [14, 15].

Recent x-ray absorption magnetic circular dichroism (XAS-MCD) experiments of several magnetic systems using the integrated transition metal core-to-core $2p$-$3s$-scattering show that spin-selective core hole screening is substantial [11] and is also important for resonant inelastic x-ray scattering (RIXS). The core hole screening is due to spin-flip processes [4] that we denote by $L_3$-$L'_3M_{4,5}$ ($L_2$-$L'_2M_{4,5}$) for excitation at the $L_3$($L_2$)-resonances. Core hole spin-flips between exchange split $2p_{3/2}$($2p_{1/2}$)-sublevels produce low energy electron/hole pairs that only occur in metals due to their lack of a bandgap. In metals, in contrast to insulators, the energy gain of such spin flips (on the order of some 0.1 eV) can be transferred to low energy electron/hole pairs close to the Fermi level thus increasing the number of core holes with spin of lowest energy. Although this could explain the reduction in the dichroic assymmetry one may ask whether lattice relaxations of the core hole via phonons and magnons are of significance too.

Consider first MnO which is an antiferromagnetic insulator with a ground state close to the ionic low spin $3d^5$-configuration [16, 17]. In principle, the simplicity of the $3d^5$-configuration (one half of the $3d$-band is filled with majority spin electrons, the other half is empty) and the large magnetic moment would make this system an ideal insulating magnetic compound in which metallic core hole screening is quenched. However, due to the superexchange mechanism, pure MnO is an antiferromagnet for which dichroic effects cancel. On the other hand, in Mn-*ferrites* the Mn-spins are *ferromagnetically* aligned and offer a close approximation to a MnO-sublattice. Thus the Mn-ferrite, as a magnetic insulator, offers an ideal system to study MCD in x-ray emission in the absence of core hole spin-flip processes, which are present in itinerant systems. Moreover, many interesting materials of scientific and technological importance today have magnetic properties of localized nature and thus it is timely to extend the scope of MCD in RIXS to this kind of materials.

In this Letter, we investigate magnetic circular dichroism at the Mn $L$-edge of $Mn_{0.6}Zn_{0.4}Fe_2O_4$ using *resonant* x-ray emission spectroscopy excited with circularly polarized x-rays with energies at the Mn $2p$ resonances. We observe that the x-ray spectra are dominated by $dd$-excitations and have no resemblance with spin-resolved density of states. Moreover, the $dd$-excitations show large dichroism at certain excitation energies, well exceeding that found in metallic systems. However, comparison to atomic crystal-field multiplet calculations shows that some other process reduces the dichroic asymmetry in this insulating magnetic system. We discuss a depolarization mechanism due to magnon and phonon coupling to the core hole excited state.

The experiments were performed at the helical undulator beamline ID12B at the European Synchrotron Radiation Facility (ESRF) in Grenoble, France [21, 22]. This beamline consists of a Dragon-like spherical grating monochromator producing 83% circularly polarized x-rays. The XAS spectra were measured in the total electron yield mode. The Rowland-type x-ray emission spectrometer [23, 24] had a 40 μm entrance slit and a spherical grating with 1800 lines/mm in the first order of diffraction, resulting in energy resolutions of 0.6 eV and 0.9 eV for XAS and RIXS, respectively. The incidence angle of the photon beam was $17^o$ and the optical axis of the spectrometer was adjusted to the surface normal thus eliminating dichroic self-absorption effects. The Mn-Zn ferrite sample was a grown single crystal thin film of $5 \times 5$ mm$^2$ surface area [25] and the measurements were made at room temperature. It was magnetized by using two Nd-Fe-B permanent magnets situated directly behind the sample with a magnetic field strength of 0.2 T.

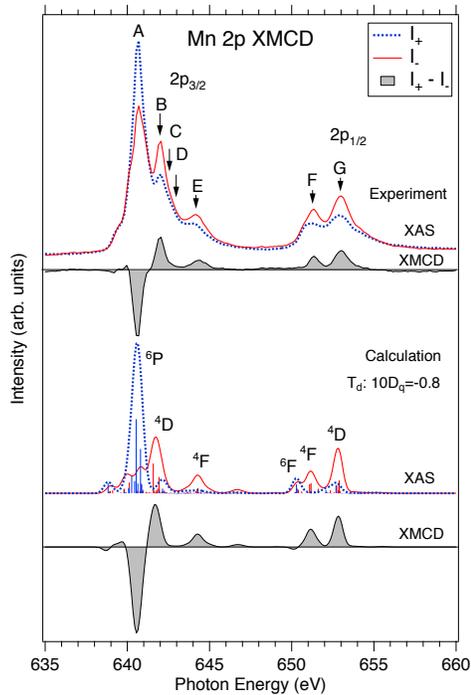

**Figure 1:** (Color online) Top, MCD in the Mn $2p$ x-ray absorption of $Mn_{0.6}Zn_{0.4}Fe_2O_4$ (top). $I_+$ represents the XAS for the magnetization parallel to the photon-spin and $I_-$ represents the antiparallel. The excitation energies used for the dichroic RIXS measurements in Fig. 2 are shown by the arrows and denoted by the letters A-G. Bottom, calculated $Mn^{2+}$ XAS and difference spectra in the tetrahedral ($T_d$) symmetry.

Figure 1 (top) shows the measured dichroic Mn $2p_{3/2,1/2}$ XAS spectra for magnetization parallel ($I_+$, solid curve) and antiparallel ($I_-$, dashed curve) to the photon spin. The filled curve is the MCD difference XAS spectrum. Calculated magnetic $Mn^{2+}$ XAS spectra for the parallel and antiparallel magnetizations are shown in the lower panel. The calculated spectra have been shifted by -2.05 eV to coincide with the experiment. The $2p_{3/2,1/2}$ peaks in the MCD spectra at 640.5 eV and 651 eV are split by approximately 11 eV by the spin-orbit interaction. The fine structures of the $2p_{3/2}$ and $2p_{1/2}$ groups consists of the crystal-field split $2p^5 3d^6$ configuration. For clarity, the main final states in spherical symmetry are indicated. The $2p_{3/2}$ and $2p_{1/2}$ thresholds are dominated by sextuplets while quadruplets dominate above. The calculated results are in good overall agreement with the experimental results although charge-transfer effects are not included. We find that the $2p$ absorption spectra are typical for divalent Mn in tetrahedral ($T_d$) symmetry and the calculations indicate that the spectra are strongly influenced by the relatively weak crystal-field interaction (optimized to -0.8 eV) between the $3d^5$ ions. The XAS spectra are dominated by strong multiplet effects due to Coulomb and exchange interactions between the $2p$ core holes and the $3d$ electrons. Note that the experimental MCD difference signal is large and that the $2p_{3/2}$ and $2p_{1/2}$ MCD peaks are opposite to each other as also predicted by our ionic model calculations. This is also the case for other Mn doped ferrites [31, 30], where the ionic model predicts a dominant single spin-down MCD peak accompanied by a weak low-energy pre-peak at 639 eV. The success of the crystal-field multiplet theory shows that the majority of the $Mn^{2+}$ ions indeed occupy the $T_d$ sites. The relative amplitudes of the calculated MCD peaks are sensitive to both the symmetry and the superexchange field.

X-ray emission (leaving a valence excitation in the final state) excited *resonantly* e.g. at the $3d$-transition metal $L$-edge of materials with localized states has been shown to be very sensitive to excitation energy. This has been explained by describing resonant x-ray emission as a *scattering process* involving two dipole transitions where energy transferred from the photon to the atom (i.e. *inelastic* scattering, thus called RIXS) is reflected as spectral weight at a corresponding energy from the elastic peak. Selection rules and large transition probabilities lead to the domination of crystal field excited final states in RIXS spectra. Divalent model calculations for RIXS of MnO in octahedral ($O_h$) symmetry have shown to be very successful in reproducing the observed excitation energies and transition intensities of *dd*- and (metal-to-ligand) charge-transfer excitations [18]. In the Mn-Zn ferrite $Mn_{0.6}Zn_{0.4}Fe_2O_4$, the magnetic $Mn^{2+}$ and $Fe^{3+}$ ions are both in the $3d^5$ state where the $Mn^{2+}$ $T_d$ and mixed valent $O_h$ $Fe^{2,3+}$ magnetic moments are antiparallel to each other [19]. However, quantitative information about the dichroism of the $Mn^{2+}$ ions can be obtained since Mn- and Fe-RIXS spectra are energetically well separated by their core electron binding energies.

Figure 2 shows the MCD in a set of experimental RIXS spectra plotted on a photon energy loss scale, with excitation energies denoted by A-G from 640.75 eV up to 652.9 eV. We observe strong energy dependent dichroism in the Mn $L_{2,3}$ RIXS-spectra of $Mn_{0.6}Zn_{0.4}Fe_2O_4$ and compare the spectra by performing crystal-field multiplet calculations using the same set of parameters as in XAS.

The RIXS spectra can be interpreted by assigning the structures to three different categories; the recombination peak, the resonating loss structures due to *dd* and charge-transfer excitations, and the normal $L_{\alpha,\beta}$ x-ray emission which is very weak at resonant energies. The elastically scattered recombination peak disperses with the excitation energy, and has a width of 0.9 eV. The recombination peak is strongest at the $2p_{3/2}$ resonance and decreases with increasing excitation energy due to the decreasing absorption cross-section. A notable success of the calculation is the relative intensity of the *dd*-excitations relative to the elastic peak. However, charge-transfer processes ($3d^6\underline{L}$ where $\underline{L}$ denotes a hole in the O 2*p* band) causing the broad structures at 10-15 eV loss energy in spectra F and G are not taken into account in the calculations.

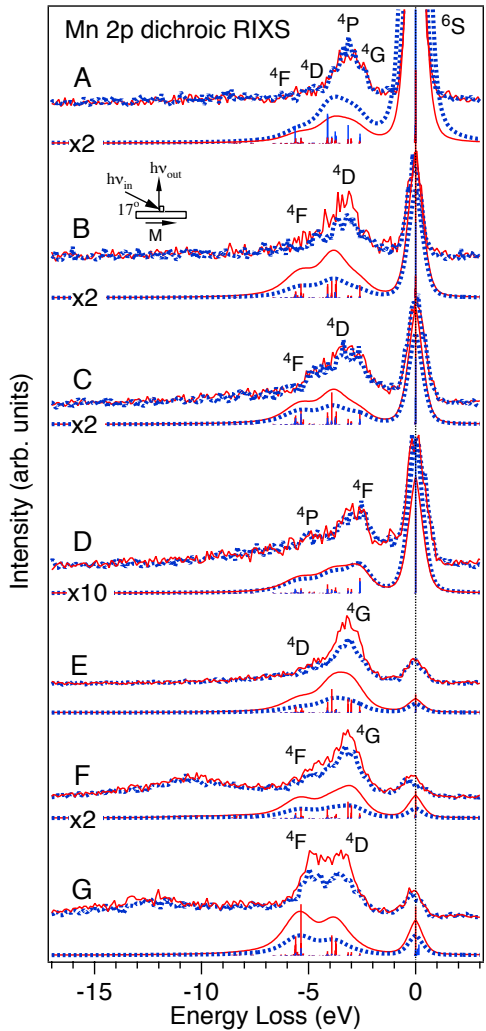

**Figure 2:** (Color online) Measured Mn $L_{2,3}$ dichroic RIXS spectra denoted A-G compared to crystal-field multiplet calculations in $T_d$ symmetry, plotted on an energy-loss scale. The measured spectra were excited at 640.75 eV, 642.1 eV, 642.4 eV, 643.0 eV, 644.4 eV, 651.1 eV and 652.9 eV photon energies, indicated by the arrows in Fig. 1. For each photon energy, the corresponding calculated spectrum is shown below.

The Mn $3d^5 \rightarrow 2p^5 3d^6 \rightarrow 3d^5$ transitions in the dichroic RIXS process were calculated as a coherent second-order optical process employing crystal-field multiplet theory in $T_d$ symmetry using the Kramers-Heisenberg formula [26]. The values of the core-level lifetime $\Gamma_i$'s used in the calculations were 0.4 eV and 0.6 eV for the $2p_{3/2}$ and $2p_{1/2}$ thresholds, respectively [27]. The Slater integrals, describing $3d-3d$ and $3d-2p$ Coulomb and superexchange interactions, and spin-orbit constants were obtained by the Hartree-Fock method [28]. The effect of the configurationally dependent hybridization was taken into account by scaling the Slater integrals $F^k(3d3d)$, $F^k(2p3d)$ and $G^k(2p3d)$ to 80 %. The ground state of the $Mn^{2+}$ ion was derived from the atomic $^6S_{5/2}$ high-spin state. The crystal-field splitting 10*Dq*, was optimized to -0.8 eV in the $T_d$ symmetry and the superexchange field to 10 meV. Calculations were made both in the $O_h$ and $T_d$ symmetries with $3d^5$ valency, with a clear preference for the $T_d$ symmetry. A direct comparison of the calculated spectra with the measured data was finally achieved by taking into account the instrumental and final state broadenings [29].

Our crystal-field multiplet calculations generally reproduce the spectral shapes of the RIXS spectra very well and the trend of the dichroic asymmetries follow the experimental ones. The final states of the *dd*-excitations are dominated by quadruplets of $^4P$, $^4D$, $^4F$ and $^4G$ symmetry. Note that the elastic recombination peak ($^6S_{5/2}$) is dichroic as a result of the dichroism in the first step of the scattering process, i.e. the absorption step. The RIXS cross-section is therefore a combination of absorption dichroism and emission dichroism, where both have been taken into account in the calculations.

Comparison between experiment and calculation reveals the important observation that, in spite of the quenching of the metallic core hole screening channel, the dichroic asymmetry at the $L_3$-resonance is reduced. Strikingly, the dichroic asymmetry is largest at the Mn $L_2$-edge (spectra F and G); this seems to be a universal effect that is observed in several metallic [3] and half-metallic [5] magnetic materials. In order to understand the difference in reduction of the dichroic asymmetry at the $L_3$- vs. $L_2$-resonance, we recall that the scattering time (also called "core hole clock" [11]), i.e. the time that the electronic system is allowed for its rearrangement, is proportional to the core hole lifetime. At the $L_2$-resonance the core hole lifetime is about 50% shorter than at the $L_3$-resonance. Hence, x-ray scattering at different resonances offers a means of studying relaxation dynamics.

Previously disregarded effects, such as spin-orbit splitting, $2p_{3/2}$ life-time broadening and the insufficient treatment of the spin quantum number, are explicitly taken care of in our calculations whereas lattice relaxation processes are not taken into account. In analogy to the metallic case ($L_3$-$L'_3 M_{4,5}$), lattice relaxations of the core-excited intermediate state can be denoted by $L_3$-$L'_3$PM and $L_2$-$L'_2$PM, where PM stands for either a phonon or a magnon. This includes nonlocal spin-flips occuring in the *core-excited* intermediate state (similarly as known for electronic screening processes [33]) as opposed to the localized spin-flip excitations in the final state discussed recently by van Veenendaal [34]. Note that the intermediate Mn $3d^6$-state is Jahn-Teller active implying that the neighboring atoms apply a torque on the 3*d*-shell at this site, as a function of its total magnetic moment. This interaction could produce magnons or optical phonons entailing a reduction of the dichroic asymmetry as a function of the available scattering time [35]. The corresponding RIXS loss energies are likely to be smaller than discernable with present instrumentation. On the other hand, using our observation we already can estimate that lattice relaxation time of a spin polarized core hole has an upper limit below 1 fs which is similar to the metallic core hole screening process.

In conclusion, we report valence RIXS dichroism at the Mn 2$p$-resonances of Mn-Zn ferrite. At resonant excitation, *dd*-excitations dominate the RIXS spectra due to the localized nature of the intermediate state and no resemblance to spin-resolved pDOS is found. We note that also 3$d$-orbitals of ferromagnetic metals have a certain degree of localization that could be of significance for their dichroic asymmetry in RIXS. The observed magnitude of the dichroic asymmetry is found to be larger in the Mn-Zn ferrite than in metallic magnetic systems, an effect of quenching of metallic core hole screening via the $L_3$-$L'_3M_{4,5}$ and $L_2$-$L'_2M_{4,5}$ decay channel. The spectral shapes and intensity trends are well reproduced by our model calculation assuming atomic-like Mn$^{2+}$-ions residing in a tetrahedral spin component of the ligand field. However, the calculated dichroic asymmetry is still larger than experiment, pointing to residual core hole relaxation mechanisms. We interpret this as existence of substantial spin-lattice interactions at the excited Mn-atom on a femtosecond time scale. Our MCD in RIXS investigation of a localized magnetic system provides an important starting point for further investigations of related ferromagnetic systems containing localized magnetic ions such as dilute magnetic semiconductors that are currently receiving strong attention for use in nanostructured hybrid materials and spintronic applications.

We acknowledge the Swedish Research Council (VR) and the Göran Gustafsson Foundation for financial support and we thank R. B. van Dover, Cornell University, for providing the sample and L. Qian and the staff at ESRF for experimental support.